\documentclass{article}
\usepackage{spconf,amsmath,graphicx,hyperref}

\usepackage{cleveref}

\usepackage{notation}

\usepackage{microtype}

\newtheorem{corollary}{Corollary}
\newtheorem{definition}{Definition}
\newtheorem{lemma}{Lemma}

\newtheorem{theorem}{Theorem}

\title{Recursive state estimation via approximate modal paths}
\name{Filip Tronarp \thanks{FT was partially supported by the Wallenberg AI, Autonomous Systems and Software Program (WASP) funded by the Knut and Alice Wallenberg Foundation}}
\address{Lund University \\ Centre for Mathematical Sciences}
\begin{document}
%
\maketitle
\begin{abstract}
In this paper, a method for recursively computing approximate modal paths is developed.
A recursive formulation of the modal path can be obtained either by backward or forward dynamic programming.
By combining both methods, a ``two-filter'' formula is demonstrated.
Both method involves a recursion over a so-called value function, which is intractable in general.
This problem is overcome by quadratic approximation of the value function in the forward dynamic programming
paradigm, resulting in both a filtering and smoothing method.
The merit of the approach is verified in a simulation experiments,
where it is shown to be on par or better than other modern algorithms.
\end{abstract}
\begin{keywords}
Nonlinear state estimation, modal paths, dynamic programming, optimization.
\end{keywords}
\section{Introduction}\label{sec:intro}
Consider the following partially observed Markov process
\begin{subequations}\label{eq:pomp}
\begin{align}
\statevar_0 &\sim \StateDensity_0(\statevar_0) \\
\statevar_{t+1} \mid \statevar_t &\sim \StateDensity_{t+1 \mid t}(\statevar_{t+1} \mid \statevar_t) \\
\obsvar_{t+1} \mid \statevar_{t+1} &\sim \ObsDensity_{t+1 \mid t+1}(\obsvar_{t+1} \mid \statevar_{t+1}),
\end{align}
\end{subequations}
where $\statevar_t$ is referred to as the (hidden) state sequence and $\obsvar_t$ are the observations.
The problem of state estimation is then to construct estimators of the state sequence based on the observations \cite{Kailath2000,Cappe2005}.
To allow for a more precise discussion introduce the following notation.
Denote the conditional density of $\statevar_{t_1:t_2}$ given $\statevar_{t_3:t_4}$ and $\obsvar_{t_5:t_6}$ by
$\StateDensity_{t_1:t_2 \mid t_3:t_4}^{t_5:t_6}$
and the conditional density of $\obsvar_{t_1:t_2}$ given $\statevar_{t_3:t_4}$ and $\obsvar_{t_5:t_6}$ by
$\ObsDensity_{t_1:t_2 \mid t_3:t_4}^{t_5:t_6}$
Further define the unnormalized conditional density of $\statevar_{t_1:t_2}$ given $\statevar_{t_3:t_4}$ and $\obsvar_{t_5:t_6}$ by
$\UnnormalizedDensity_{t_1:t_2 \mid t_3:t_4}^{t_1:t_2}  = \ObsDensity_{t_1:t_2 \mid t_1:t_2} \StateDensity_{t_1:t_2 \mid t_3:t_4}$,
e.g. $\UnnormalizedDensity_{t\mid t-1}^t = \ObsDensity_{t\mid t}\StateDensity_{t\mid t-1}$.
The goal is then to find efficient algorithms constrcut the a posteriori density over the path $\statevar_{0:T}$ given the observations $\obsvar_{0:T}$ (or various marginals), namely
\begin{equation}
\StateDensity_{0:T}^{0:T} \propto \UnnormalizedDensity_{0:T}^{0:T} =  \ObsDensity_{0:T\mid 0:T}\StateDensity_{0:T}^{0:T}.
\end{equation}
This can be done exactly only in some cases, such as when~\eqref{eq:pomp} is a linear Gaussian model~\cite{Kalman1960,RauchTungStriebel1965}.
In general approximation is required of which a popular method is the Gaussian approximation leading to algorithms such as the extended Kalman filter and smoother~\cite{Kailath2000},
cubature filters and smoothers~\cite{Julier1995,Wu2006}, the statistical linear regression method~\cite{Lefebvre2002}, and its iterative variants~\cite{Garcia2015,Garcia2016,Tronarp2018a}.
Another way to obtain Gaussian approximations is via variational inference~\cite{Courts2021a}.

Another approach to state estimation is to attempt to compute the \emph{modal path} of $\StateDensity_{0:T}^{0:T}$,
i.e. the maximum a posteriori estimate of $\statevar_{0:T}$,
leading to an optimization problem. If $\StateDensity_{t+1\mid t}$ and $\ObsDensity_{t\mid t}$ have nonlinear means and Gaussian errors,
then this problem can be solved with Gauss--Newton type methods~\cite{Bell1993,Bell1994,Skoglund2015}, Laplace-type approximations \cite{Koyama2010,Quang2015},
and in the regularized and constrained settings by augmented Lagrangians \cite{Aravkin2017,Gao2019a,Gao2019b,Gao2020a}.

\subsection{Contribution}
In this article, an approach to approximate computation of modal paths is derived based on dynamic programming \cite{Bellman1966}.
Dynamic programming has previously been used to derive the Viterbi algorithm \cite{Viterbi1967} and to derive the Kalman filter and
Rauch--Tung--Striebel smoother \cite{Larson1966}.

In order to keep the article self-contained, detailed derivations of the modal path recursions are given.
Both using backward and forward dynamic programming, which respectively results in:
\begin{enumerate}
\item Backward recursion for a value function and then a forward recursion for constructing the modal path.
\item Forward recursion for a value function and then a backward recursion for constructing the modal path.
\end{enumerate}
Additionally, a ``two-filter formula'' is derived by combining both methods.
The value function recursions are in general intractable.
Therefore, a quadratic approximation is suggested in combination with the forward dynamic programming approach,
which results in a filter-like recursion.
Similar ideas have previously been employed in optimal control under the rubric \emph{differentiable dynamic programming} \cite{DePantoja1988},
where backward dynamic programming has been favoured.
To the author's knowledge these ideas have not been employed in the context of state estimation.

The rest of this paper is organized as follows.
The recursive theory of modal paths is outlined in \cref{sec:recursive-modal-path-theory}.
From this, a serviceable approximate algorithm is constructed based on quadratic expansions in \cref{sec:approximate-modal-paths}.
Simulation experiments are carried out in \cref{sec:experiments} and conclusions are drawn in \cref{sec:conclusion}.

\section{Recursive computation of modal paths}\label{sec:recursive-modal-path-theory}
In this section, a recursive formulation of the modal path computation is formulated in the abstract setting.
Let $\gamma$ denote the logarithm of the unnormalized conditional state density
\begin{equation}
\gamma_{t_1:t_2 \mid t_3:t_4}^{t_1:t_2} = \log\UnnormalizedDensity_{t_1:t_2 \mid t_3:t_4}^{t_1:t_2}.
\end{equation}
Then the modal path problem is given by
\begin{equation}
\label{eq:modal-path-definition}
\begin{split}
\statevar_{0:T}^{0:T} &= \argmax[\statevar_{0:T}] \gamma_{0:T}^{0:T}(\statevar_{0:T}) \\
&= \argmax[\statevar_{0:T}] \Big[ \gamma_0^0(\statevar_0) + \sum_{t=1}^T \gamma_{t \mid t-1}^t(\statevar_t \mid \statevar_{t-1}) \Big].
\end{split}
\end{equation}
In \cref{sec:backward-recursion}, recursive solutions to~\eqref{eq:modal-path-definition} are derived on based on backward dynamic programming \cite{Bellman1966}.
In \cref{sec:forward-recursion} another recursion is derived based on \emph{forward} dynamic programming,
which is more useful in the sense that it gives a filter-like recursion for the terminal modal states $\statevar_t^{0:t}$.
Finally, for completeness the two methods are combined in \cref{sec:two-filter}  to give a ``two-filter formula'' for $\statevar_t^{0:T}$.

\subsection{Backward dynamic programming}\label{sec:backward-recursion}
In backward dynamic programming, the maximization problem is simplified by defining a backward value function.
In the context of~\eqref{eq:modal-path-definition}, it is the maxima of the modal path problem for the sub-paths $\statevar_{t+1:T}$ conditionally on $\statevar_t$.
\begin{definition}[Backward value function for modal paths]\label{def:backward-value-function-modal-paths}
Define the backward value function for modal paths at time $T$ by $\BackwardValueFunction_{T}(x_T) = 0$ and for time $t$ with $0 \leq t < T$  the backward value function is given by
\begin{equation}\label{eq:backward-value-function-modal-paths}
\BackwardValueFunction_t(\statevar_t) = \max_{\statevar_{t+1:T}}\, \gamma_{t+1:T\mid t}^{t+1:T}(\statevar_{t+1:T} \mid \statevar_t).
\end{equation}
\end{definition}
The rationale behind this definition is that the optimization problem of~\eqref{eq:modal-path-definition} can formally be reduced to an optimization problem over $t+1$ state variables for $t \geq 0$.
More specifically, by using the Markov property and the definition of backward value function the maximization problem  of~\eqref{eq:modal-path-definition} is for any $0 \leq t \leq T$ given by
\begin{equation*}
\begin{split}
&\max_{\statevar_{0:T}}\, \gamma_{0:T}^{0:T}(\statevar_{0:T})
= \max_{\statevar_{0:T}}\, \big[ \gamma_{0:t}^{0:t}(\statevar_{0:t}) +  \gamma_{t+1:T\mid t}^{t+1:T}(\statevar_{t+1:T} \mid \statevar_t) \big] \\
&\quad = \max_{\statevar_{0:t}}\,\max_{\statevar_{t+1:T}}\, \big[ \gamma_{0:t}^{0:t}(\statevar_{0:t}) +  \gamma_{t+1:T\mid t}^{t+1:T}(\statevar_{t+1:T} \mid \statevar_t) \big] \\
&\quad = \max_{\statevar_{0:t}}\,\Big[ \gamma_{0:t}^{0:t}(\statevar_{0:t}) +  \BackwardValueFunction_t(\statevar_t) \Big].
\end{split}
\end{equation*}
The preceeding calculation establishes the following lemma.
\begin{lemma}\label{lem:forward-mode-decomposition}
The maximum of $\gamma_{0:T}^{0:T}$ is for $0 \leq t \leq T$ given by
\begin{equation}
\max_{\statevar_{0:T}}\, \gamma_{0:T}^{0:T}(\statevar_{0:T})
= \max_{\statevar_{0:t}}\,\Big[ \gamma_{0:t}^{0:t}(\statevar_{0:t}) +  \BackwardValueFunction_t(\statevar_t) \Big].
\end{equation}
\end{lemma}
Given the backward value function, $\BackwardValueFunction_t$,
\cref{lem:forward-mode-decomposition} provides a means to construct the entire modal path, $x_{0:T}^{0:T}$, by a forward recursion.
Namely, set $t = 0$ in \cref{lem:forward-mode-decomposition} then the initial condition of the modal path is given by
\begin{equation}
\statevar_0^{0:T} = \argmax[\statevar_0]  \big[ \gamma_0^0(\statevar_0) +  \BackwardValueFunction_0(\statevar_0) \big].
\end{equation}
Furthermore, the Markov property and \cref{lem:forward-mode-decomposition} then gives
\begin{equation*}
\begin{split}
&\max_{\statevar_{0:T}}\, \gamma_{0:T}^{0:T}(\statevar_{0:T})
= \max_{\statevar_{0:t}}\,\Big[ \gamma_{0:t}^{0:t}(\statevar_{0:t}) +  \BackwardValueFunction_t(\statevar_t) \Big] \\
&\quad = \max_{\statevar_{0:t}}\,\Big[  \gamma_{0:t-1}^{0:t-1}(\statevar_{0:t-1}) + \gamma_{t \mid t - 1}^t(\statevar_t \mid \statevar_{t-1})
+  \BackwardValueFunction_t(\statevar_t) \Big] \\
&=  \gamma_{0:t-1}^{0:t-1}(\statevar_{0:t-1}^{0:T}) + \max_{\statevar_t}\,
\big[\gamma_{t \mid t - 1}^t(\statevar_t \mid \statevar_{t-1}^{0:T}) +  \BackwardValueFunction_t(\statevar_t) \big],
\end{split}
\end{equation*}
where the maximization over $\statevar_{0:t}$ was split into first maximization over $\statevar_{0:t-1}$ and then a maximization over $\statevar_t$.
\Cref{thm:modal-path-forward-recursion} has now been proven.
\begin{theorem}[Forward recursion for modal path]\label{thm:modal-path-forward-recursion}
Given the backward value function $\BackwardValueFunction_t$ for $0 \leq t \leq T$,
the modal trajectory is recursively given by
\begin{subequations}
\begin{align*}
\statevar_0^{0:T} &= \argmax[\statevar_0] \big[ \gamma_0^0(\statevar_0) +  \BackwardValueFunction_0(\statevar_0) \big] \\
\statevar_t^{0:T} &= \argmax[\statevar_t] \big[ \gamma_{t \mid t - 1}^t(\statevar_t \mid \statevar_{t-1}^{0:T}) + \BackwardValueFunction_t(\statevar_t) \big],
\, 1 \leq t \leq T.
\end{align*}
\end{subequations}
\end{theorem}
The backward value function needs to be obtained in order to make \cref{thm:modal-path-forward-recursion} useful in practice.
It remains to show that it satisfies a certain backward recursion.
This follows immediately from the \cref{def:backward-value-function-modal-paths} and the Markov property
\begin{equation*}
\begin{split}
&\BackwardValueFunction_{t-1}(\statevar_{t-1})
= \max_{\statevar_{t:T}}\, \gamma_{t:T\mid t-1}^{t:T}(\statevar_{t:T} \mid \statevar_{t-1}) \\
&\quad = \max_{\statevar_{t:T}}\, \Big[ \gamma_{t\mid t-1}^t(\statevar_t \mid \statevar_{t-1}) +  \gamma_{t+1:T\mid t}^{t+1:T}(\statevar_{t+1:T} \mid \statevar_t) \Big] \\
&\quad = \max_{\statevar_t} \Big[ \gamma_{t\mid t-1}^t(\statevar_t \mid \statevar_{t-1}) + \BackwardValueFunction_t(x_t) \Big],
\end{split}
\end{equation*}
and since $\BackwardValueFunction_T(x_T) = 0$ by definition this recursion hold for any $1 \leq t \leq T$.
\Cref{thm:modal-path-backward-value-function-recursion} has been verified.
\begin{theorem}[Backward value function recursion]\label{thm:modal-path-backward-value-function-recursion}
The backward value function for the modal path satisfies for $1 \leq t \leq T$ the following recursion
\begin{equation}
\BackwardValueFunction_{t-1}(\statevar_{t-1}) =  \max_{\statevar_t} \Big[ \gamma_{t\mid t-1}^t(\statevar_t \mid \statevar_{t-1}) + \BackwardValueFunction_t(x_t) \Big],
\end{equation}
with terminal condition $\BackwardValueFunction_T(x_T) = 0$.
\end{theorem}
The modal path may now formally be obtained by first computing the backward value function via \cref{thm:modal-path-backward-value-function-recursion} and
then construcing the modal path via \cref{thm:modal-path-forward-recursion}.

\subsection{Forward dynamic programming}\label{sec:forward-recursion}
In forward dynamic programming, maximization problem is simplified by defining a forward value function.
In the context of the optimization problem~\eqref{eq:modal-path-definition},
it is the maxima of the modal path from for sub-paths $\statevar_{0:t-1}$ conditionally on $\statevar_t$.
\begin{definition}[Forward value function for modal paths]\label{def:forward-value-function-modal-paths}
Define the forward value function for modal paths by
\begin{equation}
\ForwardValueFunction_t(\statevar_t) = \max_{\statevar_{0:t-1}} \gamma_{0:t}^{0:t}(\statevar_{0:t}), \quad 1 \leq t \leq T,
\end{equation}
and define the initial condition by $\ForwardValueFunction_0(\statevar_0) = \gamma_0^0(\statevar_0)$.
\end{definition}
Similarly to the backward value function, the forward value function can be used to reduce the optimization problem of \eqref{eq:modal-path-definition} into a problem with fewer state variables.
This is again accomplished by exploiting the Markov property, namely
\begin{equation*}
\begin{split}
&\max_{\statevar_{0:T}}\, \gamma_{0:T}^{0:T}(\statevar_{0:T})
= \max_{\statevar_{0:T}}\, \big[ \gamma_{0:t}^{0:t}(\statevar_{0:t}) +  \gamma_{t+1:T\mid t}^{t+1:T}(\statevar_{t+1:T} \mid \statevar_t) \big] \\
&\quad = \max_{\statevar_{0:t-1}}\,\max_{\statevar_{t:T}}\, \big[ \gamma_{0:t}^{0:t}(\statevar_{0:t}) +  \gamma_{t+1:T\mid t}^{t+1:T}(\statevar_{t+1:T} \mid \statevar_t) \big] \\
&= \max_{\statevar_{t:T}}\,   \big[ \ForwardValueFunction_t(\statevar_t) +  \gamma_{t+1:T\mid t}^{t+1:T}(\statevar_{t+1:T} \mid \statevar_t) \big]
\end{split}
\end{equation*}
This demonstrates \cref{lem:backward-mode-decomposition}.
\begin{lemma}
\label{lem:backward-mode-decomposition}
The maximum of $\gamma_{0:T}^{0:T}$ is for $0 \leq t < T$ given by
\begin{equation*}
\max_{\statevar_{0:T}}\, \gamma_{0:T}^{0:T}(\statevar_{0:T}) =
\max_{\statevar_{t:T}}\,   \big[ \ForwardValueFunction_t(\statevar_t) +  \gamma_{t+1:T\mid t}^{t+1:T}(\statevar_{t+1:T} \mid \statevar_t) \big]
\end{equation*}
\end{lemma}
The terminal value of the modal path may be obtained directly from the forward value function by setting $t = T$ in \cref{lem:backward-mode-decomposition}
\begin{equation}
\statevar_T^{0:T} =\argmax[\statevar_t]\, \ForwardValueFunction_T(x_T).
\end{equation}
On the other hand, for $t < T$ \cref{lem:backward-mode-decomposition} gives
\begin{equation*}
\max_{\statevar_{0:T}}\, \gamma_{0:T}^{0:T}(\statevar_{0:T}) =
\max_{\statevar_{t:T}}\,   \Big[ \ForwardValueFunction_t(\statevar_t) +  \gamma_{t+1:T\mid t}^{t+1:T}(\statevar_{t+1:T} \mid \statevar_t) \Big]
\end{equation*}
Therefore, the modal path at time $t$ is given  by
\begin{equation*}
\statevar_t^{0:T} = \argmax[\statevar_t]  \Big[ \ForwardValueFunction_t(\statevar_t) +  \gamma_{t+1:T\mid t}^{t+1:T}(\statevar_{t+1:T}^{0:T} \mid \statevar_t) \Big],
\end{equation*}
but due the Markov property again, this simplifies to
\begin{equation}
\statevar_t^{0:T} = \argmax[\statevar_t]  \Big[ \ForwardValueFunction_t(\statevar_t) +  \gamma_{t+1\mid t}^{t+1}(\statevar_{t+1}^{0:T} \mid \statevar_t) \Big].
\end{equation}
A backward recursion for the modal path has now been derived, as stated in \cref{thm:modal-path-backward-recursion}.
\begin{theorem}[Backward recursion for modal path]
\label{thm:modal-path-backward-recursion}
Given the forward value function $\ForwardValueFunction_t$ for $0 \leq t \leq T$,
the modal path is recursively given by
\begin{equation}
\statevar_t^{0:T} = \argmax[\statevar_t]  \big[ \ForwardValueFunction_t(\statevar_t) +  \gamma_{t+1\mid t}^{t+1}(\statevar_{t+1}^{0:T} \mid \statevar_t) \big],
\end{equation}
where the terminal value is determined by
\begin{equation}
\statevar_T^{0:T} = \argmax[\statevar_T] \, \ForwardValueFunction_T(x_T).
\end{equation}
\end{theorem}
It remains to obtain a recursion for the forward value function.
By \cref{def:forward-value-function-modal-paths} and the Markov property,
the forward value function is at time $t+1$  given by
\begin{equation}
\begin{split}
\ForwardValueFunction_{t+1}(\statevar_{t+1}) &= \max_{\statevar_{0:t}} \gamma_{0:t+1}^{0:t+1}(\statevar_{0:t+1})  \\
&= \max_{\statevar_{0:t}}
\big[ \gamma_{0:t}^{0:t}(\statevar_{0:t}) + \gamma_{t+1\mid t}^{t+1}(\statevar_{t+1}  \mid \statevar_t ) \big] \\
&= \max_{\statevar_t}
\big[ \ForwardValueFunction_t(\statevar_t) + \gamma_{t+1\mid t}^{t+1}(\statevar_{t+1}  \mid \statevar_t ) \big].
\end{split}
\end{equation}
The result is summarized in the following theorem.
\begin{theorem}[Forward value function recursion]\label{thm:modal-path-forward-value-function-recursion}
The forward value function for the modal path satisfies the following recursion
\begin{equation}
\ForwardValueFunction_{t+1}(\statevar_{t+1}) =
\max_{\statevar_t} \big[ \ForwardValueFunction_t(\statevar_t) + \gamma_{t+1\mid t}^{t+1}(\statevar_{t+1}  \mid \statevar_t ) \big].
\end{equation}
The initial condition is determined by \cref{def:forward-value-function-modal-paths}, namely
\begin{equation}
\ForwardValueFunction_0(\statevar_0)
= \gamma_0^0(\statevar_0)
\end{equation}
\end{theorem}
Since $T$ is arbitrary in \cref{thm:modal-path-backward-recursion} it can be seen that the state at time $t$ of the modal path associated with $\StateDensity_{0:t}^{0:t}$ is
the maximizer of the forward value function.
\begin{corollary}\label{cor:map-filter}
The terminal value of the modal path of $\StateDensity_{0:t}^{0:t}$ is given by
\begin{equation}
\statevar_t^{0:t} = \argmax[\statevar_t] \ForwardValueFunction_t(\statevar_t).
\end{equation}
\end{corollary}
This result can be combined with \cref{thm:modal-path-forward-value-function-recursion} to recursively compute the mode of the filtering densities on-line.

\subsection{A ``two-filter'' formula}\label{sec:two-filter}
The backward and forward dynamic programming methods may be combined to get an expression for $\statevar_t^{0:T}$ in terms of both the backward and the forward value function.
This is the dynamic programming analogue of the two-filter formula.
From \cref{lem:backward-mode-decomposition} and \cref{def:backward-value-function-modal-paths} $\statevar_t^{0:T}$ is given by
\begin{equation*}
\begin{split}
\statevar_t^{0:T} &= \argmax[\statevar_t] \underset{x_{t+1:T}}{\max}\,   \big[ \ForwardValueFunction_t(\statevar_t) +  \gamma_{t+1:T\mid t}^{t+1:T}(\statevar_{t+1:T} \mid \statevar_t) \big] \\
&= \argmax[\statevar_t] \big[ \ForwardValueFunction_t(\statevar_t) +  \BackwardValueFunction_t(\statevar_t) \big],
\end{split}
\end{equation*}
which proves \cref{cor:modal-tw-filter}.
\begin{corollary}\label{cor:modal-tw-filter}
The state of the modal path at time $t$ is given by
\begin{equation}
\statevar_t^{0:T} =  \argmax[\statevar_t]  \big[ \ForwardValueFunction_t(\statevar_t) +  \BackwardValueFunction_t(\statevar_t) \big].
\end{equation}
\end{corollary}

\section{Approximate modal paths}\label{sec:approximate-modal-paths}
The recursions for the modal path and value functions established in \cref{sec:backward-recursion} and \cref{sec:forward-recursion} are of course intractable in general.
In this section, approximate methods for the forward dynamic programming approach are suggested.
This is based on the following quadratic approximation
\begin{equation}\label{eq:approximate-forward-value-function}
\widehat{\ForwardValueFunction_t}(\statevar_t) = \log \kappa_t
- \frac{1}{2}(\statevar_t -  \StateMean_t)^* \StateCov_t^{-1} (\statevar_t -  \StateMean_t)
\end{equation}
Now define the following objective function
\begin{equation}
v_{t:t+1}(\statevar_{t+1}, \statevar_t) = \widehat{\ForwardValueFunction_t}(\statevar_t) + \gamma_{t+1\mid t}^{t+1}(\statevar_{t+1}  \mid \statevar_t),
\end{equation}
then assuming that $\widehat{\ForwardValueFunction_t}(\statevar_t) = \ForwardValueFunction_t(\statevar_t)$ gives the forward value function at time $t+1$ as (c.f. \cref{thm:modal-path-forward-value-function-recursion})
\begin{equation}\label{eq:forward-value-function-recursion-2}
\ForwardValueFunction_{t+1}(\statevar_{t+1}) = \max_{\statevar_t}  v_{t:t+1}(\statevar_{t+1}, \statevar_t).
\end{equation}
This problem is still untractable in general and the last step to ``close'' the recursion is to replace $v_{t:t+1}$ by a quadratic approximation
according to
\begin{equation*}
\begin{split}
&v_{t:t+1}(\statevar_{t+1}, \statevar_t) \approx \hat{v}_{t:t+1}(\statevar_{t+1}, \statevar_t) =  \widehat{\ForwardValueFunction_{t+1}}(\statevar_{t+1})  \\
&\quad -\frac{1}{2} (\statevar_t - \StateCondMean_{t\mid t+1}(\statevar_{t+1}))^* \StateCondCov_{t\mid t+1}^{-1}
(\statevar_t - \StateCondMean_{t\mid t+1}(\statevar_{t+1})),
\end{split}
\end{equation*}
where $\widehat{\ForwardValueFunction_{t+1}}$ is of the same form as~\cref{eq:approximate-forward-value-function} and $\StateCondMean_{t\mid t+1}$ is an affine function, say
\begin{equation}
\StateCondMean_{t\mid t+1}(\statevar_t) = \TransitionMatrix_{t, t+1} \statevar_{t+1} + \StateControl_{t+1, t}.
\end{equation}
Such a quadratic approximation can be obtained by a second order Taylor expansions around the mode of $v_{t:t+1}$,
or by a Gauss--Newton linearizartion when applicable, or a combination of the two (when applicable).
In any case, the mode of $\hat{v}_{t:t+1}$ with respect to $\statevar_t$ is then $\StateCondMean_{t\mid t+1}(\statevar_{t+1})$,
hence replacing $v_{t:t+1}$  in~\eqref{eq:forward-value-function-recursion-2} by $\hat{v}_{t:t+1}$ gives
\begin{equation}
\ForwardValueFunction_{t+1}(\statevar_{t+1}) \approx \max_{\statevar_t}  \hat{v}_{t:t+1}(\statevar_{t+1}, \statevar_t) = \widehat{\ForwardValueFunction_{t+1}}(\statevar_{t+1}).
\end{equation}
Lastly, a backward recursion for the approximate modal path is obtained by (c.f. \cref{thm:modal-path-backward-recursion})
\begin{equation}
\hat{\statevar}_t^{0:T} = \argmax[\statevar_t]  \hat{v}_{t:t+1}(\hat{\statevar}_{t+1}^{0:T}, \statevar_t)  = \StateCondMean_{t\mid t+1}(\hat{\statevar}_{t+1}^{0:T}),
\end{equation}
and the terminal condition is the maximizer of $\widehat{\ForwardValueFunction_T}$, namely $\StateMean_T$.

\section{Experiments}\label{sec:experiments}
The approximate modal path algorithm (AMP) is tested on the stochastic Ricker map, which is given by
\begin{subequations}
\begin{align}
\StateDensity_0(\statevar_0) &= \gaussian(\log 7, 0.1^2) \\
\StateCondMean_{t\mid t-1}(\statevar_{t-1}) &= \statevar_{t-1} - e^{\statevar_{t-1}} +\log 44.7 \\
\StateDensity_{t\mid t-1}(\statevar_t \mid \statevar_{t-1}) &= \mathcal{N}(\statevar_t; \StateCondMean_{t\mid t-1}(\statevar_{t-1}), 0.3^2) \\
\ObsDensity_{t\mid t}(\statevar_t) &= \mathrm{Po}(2e^{\statevar_t}).
\end{align}
\end{subequations}
This model is \emph{highly} nonlinear and has previously been used as a benchmark by for example \cite{Tronarp2018a}.
The method is compared to the Kalman Laplace filter (KLF) proposed by \cite{Quang2015} in addition to the iterated statistical linear regression method (IPLF) proposed by \cite{Tronarp2018a}.
The IPLF uses first order Taylor linearizations.
A quadratic approximation of $v_{t:t+1}$ is obtained by a Gauss-Newton linearization of $\log \StateDensity_{t\mid t-1}$
and a second order Taylor expansion of $\log \ObsDensity_{t\mid t}$ around the mode of $v_{t:t+1}$.

\begin{figure}
\centering
\includegraphics{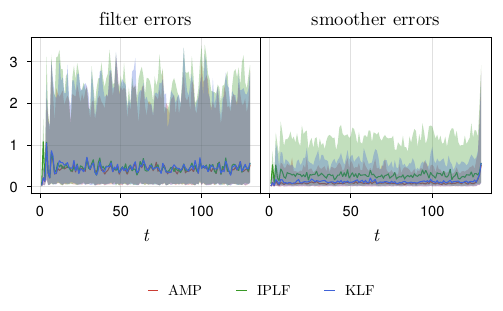}
\caption{
The absolute estimation errors for the filters (first row) and smootherts (second row).
The solid lines are the median over all trials and the shaded regions correspond to the interval between the
10\% and 90\% quantiles.
}
\label{fig:ricker}
\end{figure}

The system is simulated $T = 2^7$ time-steps and the methods are compared in terms of absolute error
in the filter and smoother over 100 simulations. The results are shown in \cref{fig:ricker}.

It can be seen that all filters perform similarly with KLF having a moderate advantage over IPLF and
AMP having a small advantage over KLF.
Furthermore, the median of KLF and AMP perform similarly in the smoother, with a small advantage to AMP.
On the other hand, AMP outperforms KLF to a moderate degree in terms of the 90\% quantiles of the absolute error.
This demonstrates that the performance of AMP is more reliable in this problem.

\section{Conclusion}\label{sec:conclusion}

The recursive relations have been reviewed for the modal path in partially observed Markov processes
based on both backward and forward dynamic programming.
Both methods were combined to arrive at a ``two-filter'' formula.
It is an open problem whether backward method or the ``two-filter'' formula can be leveraged
in the design of state estimation algorithms.
However, the forward method was turned into a serviceable algorithm by means of quadratic approximation.
The resulting algorithm was shown to outperform other modern algorithms in a simulation experiment.

Future work involves extending the approach to regularizerd and constrained state estimation problems as considered in \cite{Gao2019b,Gao2020a}
and to generalized state estimation \cite{Aravkin2017}.

\vfill\pagebreak

\bibliographystyle{IEEEbib}
\bibliography{refs}

\end{document}